# The use of Recommender Systems in web technology and an in-depth analysis of Cold State problem

Denis Selimi[1], Krenare PIREVA NUCI[2]

[1][2]University for Business and Technology, Faculty of Computer Science and Engineering

ds33387@ubt-uni.net, krenare.pireva@ubt-uni.net

**Abstract.** In the WWW (World Wide Web), dynamic development and spread of data has resulted a tremendous amount of information available on the Internet, yet user is unable to find relevant information in a short span of time. Consequently, a system called recommendation system developed to help users find their infromation with ease through their browsing activities. In other words, recommender systems are tools for interacting with large amount of information that provide personalized view for prioritizing items likely to be of keen for users. They have developed over the years in artificial intelligence techniques that include machine learning and data mining amongst many to mention. Furthermore, the recommendation systems have personalized on an e-commerce, on-line applications such as Amazon.com, Netflix, and Booking.com. As a result, this has inspired many researchers to extend the reach of recommendation systems into new sets of challenges and problem areas that are yet to be truly solved, primarily a problem with the case of making a recommendation to a new user that is called cold-state (i.e. cold-start) user problem where the new user might likely not yield much of information searched. Therfore, the purpose of this paper is to tackle the said cold-start problem with a few effecient methods and challenges, as well as identify and overview the current state of recommendation system as a whole.

## 1. Introduction

In this ever-changing technological world of advancement with diverse amount of data on the internet, has led the user to be overloaded with information [1]. The recommender system (RS) thus allows us to manage and effectively deal with said overload, by classifying the set of items into categorized niches that later can be recommended by users who search for it, whether be it textual, audio, image, and video in which are constantly updated.

Recommendation system nowadays, allows us to explore variety of items that may make user content from recommendations they provide – be it in the movies, music, or news. Such example amongst many, are Amazon that recommends books, Booking.com recommending accommodations for the users, Ebay known for using item-to-item based collaboration, and Netflix who recommends movies -  all that provide and benefit from recommending to their clients [2]. This recommendation is determined by variety techniques, as highlighted by many scientific studies [3]. Giving examples by some of the techniques, include an improvement over the transaction costs of finding and selecting specific items in online shopping environment, for e-commerce purposes for selling more products as a tool that searches users' preferences, and in scientific libraries whereas the RS aids in users to search beyond the specified niches. Therefore, it is indispensable to understand the concept of converting the data information into recommendation techniques within a system that will be a convenient approach for the users to not be overloaded.

The combination of RS concepts and using its techniques are a new understanding for creating a recommendation e-commerce site to predict the users' preferences on an item. The work of this topic has been achieved using data from various sources that have been critically checked and analyzed.

In this paper, a new framework based on web applications (i.e. e-commerce sites) of recommendation systems and its techniques are proposed to overcome the drawbacks associated mostly from the cold state situation that will further be defined in the problem declaration section within paper. Although the main point within this paper relies on its problem with cold state in problem declaration section, in the section two lists important the concepts of recommendation system in which it uses specific variation of algorithms and techniques, mostly being the content-based filtering and its user-based analysis, collaborative filtering technique. In the next section are included theories, terms, and substantive findings in which will help tremendously in analyzing the information concepts, i.e. content-based filtering, and by giving examples of other users' accesses of behaviors such as collaborative filtering.

### 1.1. Problem declaration

The use of recommender system in intelligent application is common nowadays; however, besides the advantages that are coming with the new technologies, there are still gaps that are phased in the very beginning of these systems. One of the main gaps that the system may face is the cold state situation, where the system is new and there is no data recorded regarding the users' profiles. Thus, the main problem of this paper is how we can escape this situation and whether the system will be able to have the intelligent part from the very beginning.

The aim of the problem declaration within recommendation system is to tackle the cold-state (i.e. cold-start) situation in web applications that tends to halt certainty in e-commerce companies or sites (e.g. Amazon, Netflix) and/or aforementioned Booking.com site. So in order to give possible treatments, in the section 5.0 will be further stated in ways to directly solidify the cold-state situation itself. There are also mentionings of techniques in the previous section that aid to define the RS.

The objectives of this paper is to: (1): Review the use of recommender systems in intelligent systems, (2) to list and compare a set of algorithms used in RS and (3) to list a few scenarios to possibly treat in continuous cold state situation from certain users.

The paper is strucutured as follows, section 2 gives the general knowledge how recommender systems work and the related work in this domain, continuing with section 3, which give an in-depth analysis of Cold State problem within initial phase of Recommender Systems.The discussion is provided in section 4, and finally concluding the paper in section 5.

## 2. General Concepts of Recommender Systems and Related work

In this section is addressed the use of algorithms and different style of techniques that help identify recommendation system itself, by initially listing the algorithms used in recommendation systems and its data mining techniques that identifies a good algorithm for RS (see Figure 1).

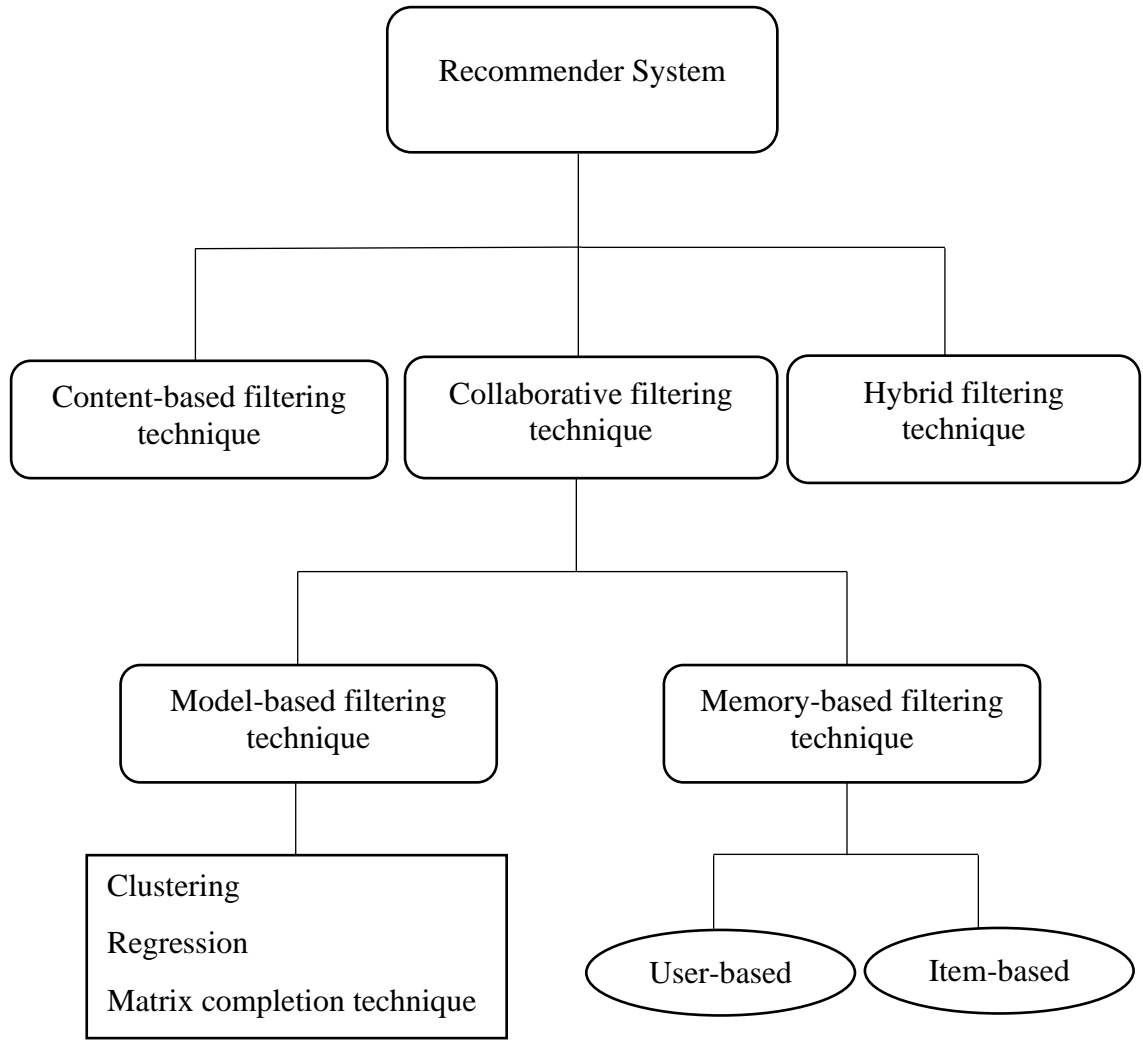

*Figure 1: Block diagram of recommendation system techniques.*

As shown in Figure 1 Recommender System use Content, Collaborative and Hybrid Filtering techniques to recommend a number of items depending in which domain we are applying them. Collaborative filtering technique will specify its in-depth relations between the user and set of items, giving us two subsections: Memory-based technique and mode-based technique within collaborative filtering. Once the filtering techniques are mentioned, the hybrid filtering has its importance to tie and conclude how the relations between CF and CBF can be combined to yield a better recommendation system.

**2.1 The algorithms used in RS**

The amount of data being generated nowadays has increased exponentially, whether being business related such as purchase transactions, to scientific domains such as satellite image datas. Nowadays, the algorithms within RS are used with data mining, in which it exhibits the following characteristics:

**High-level efficiency:** The information that it is discovered should in a high-level language which is understood by human users. Furthermore, the discovery process of said information is efficient and yields the running times for the mass sizes of databases acceptable.

**Unique results with accuracy:** Said discovered information should have an interesting, unique result according to user biases. In particular, having a unique result with accurate contents of databases portrays that patterns are novel and perhaps useful.

With that said, the data mining is crucial when it comes to recommendation system and data science branch itself, as it helps in research fields such as database management, statistics, and machine learning; therefore, there are three classes of data mining algorithms that are applied in various niches of applications, and they are as follows:

**Classification:** This type of algorithm is widely known amongst data mining algorithm and has been an ongoing study in the research field of machine learning for several decades. The main goal of classification algorithm is to classify cases into different classes that gets based on attributes among a set of objects within database. Such example in classification gets used in medical diagnosis, target niches of marketing, fraud detection, as well as the information retrieval.

**Association rule mining:** This is a method – or a procedure – that is meant to find frequent patterns. The output of this method has a set of rules where the occurrence of items in precondition implies the co-occurrence of items in precondition. They are found in various kinds of databases such as relational databases, transactions, and e-commerce sites.

**Clustering:** The clustering algorithm, or in other words called unsupervised classification, is the process of grouping physical or abstract objects in such a way, where the objects in the same group are more similar to each other than to those other groups. Hencemore, clustering analysis helps constructing a meaningful partitioning on the mass size of objects based on the characteristics. The clusting is used and will be mentioned more in the model-based collaborative technique within recommendation system. Below is a table used as an example of clustering algorithm.

Table 1: Cluster example with customers recommending books.

|            | Book 1 | Book 2 | Book 3 | Book 4 | Book 5 | Book 6 |
|------------|--------|--------|--------|--------|--------|--------|
| Customer A | X      |        |        | X      |        |        |
| Customer B |        | X      | X      |        | X      |        |
| Customer C |        | X      | X      |        |        |        |
| Customer D |        | X      |        |        |        | X      |
| Customer E | X      |        |        |        | X      |        |
| Customer F |        |        | X      |        | X      |        |

As shown in Table 1, any customer shall be classified as a member of a cluster and will receive recommendations based on preferences on the particular group. As it is shown, book 2 will be highly recommended to Customer F, as well as the book 6 will be recommended to some extent. Therefore, Customer B, C, and D form a cluster – marked in green – and Customer A with E will form another cluster towards each other.

**2.2 Content-based filtering**

Content-based filtering technique (CBF) is a domain-dependent algorithm that relies more upon the analysis of the attributes of items in order to generate predictions. It is a part of recommender system that handles with profiles of users that get created in the beginning or from the past searches depending on the niche. In its process, the engine compares the items that were already positively rated and recommends to the user accordingly. CBF uses different sets of models to find similarity between sets of documents in order to generate the right set of recommendations to the user based on its profile [4]. One of such approach is with cosine similarity that measures to find contents that matches user profile. It uses vectors of two items (wc and ~ ws) with attributes that compares in similarity function:

$$u(c,s) = \cos(\vec{w_c} \cdot \vec{w_s}) = \frac{\vec{w_c} \cdot \vec{w_s}}{\|\vec{w_c}\| \times \|\vec{w_s}\|} = \frac{\sum_{i=1}^{K} \vec{w}_{ic} \vec{w}_s}{\sqrt{\sum_{i=1}^{K} \vec{w}_{ic}^2} \sqrt{\sum_{i=1}^{K} \vec{w}_{is}^2}} \quad \ldots\ldots(1)$$

Moreover on vectors, CBF also models the relations between different documents within a bulk of items such as: Term frequency, naïve Bayes classifier, or neural networks. If the user profile changes, CBF's technique still has the potential to adjust itself within a short period of time. So the profile is often updated automatically when there is a feedback on the items' desirability that has been presented to the user. In the figure below is shown a diagram for CBF and architecture named 'State of the Art and Trends' [5].

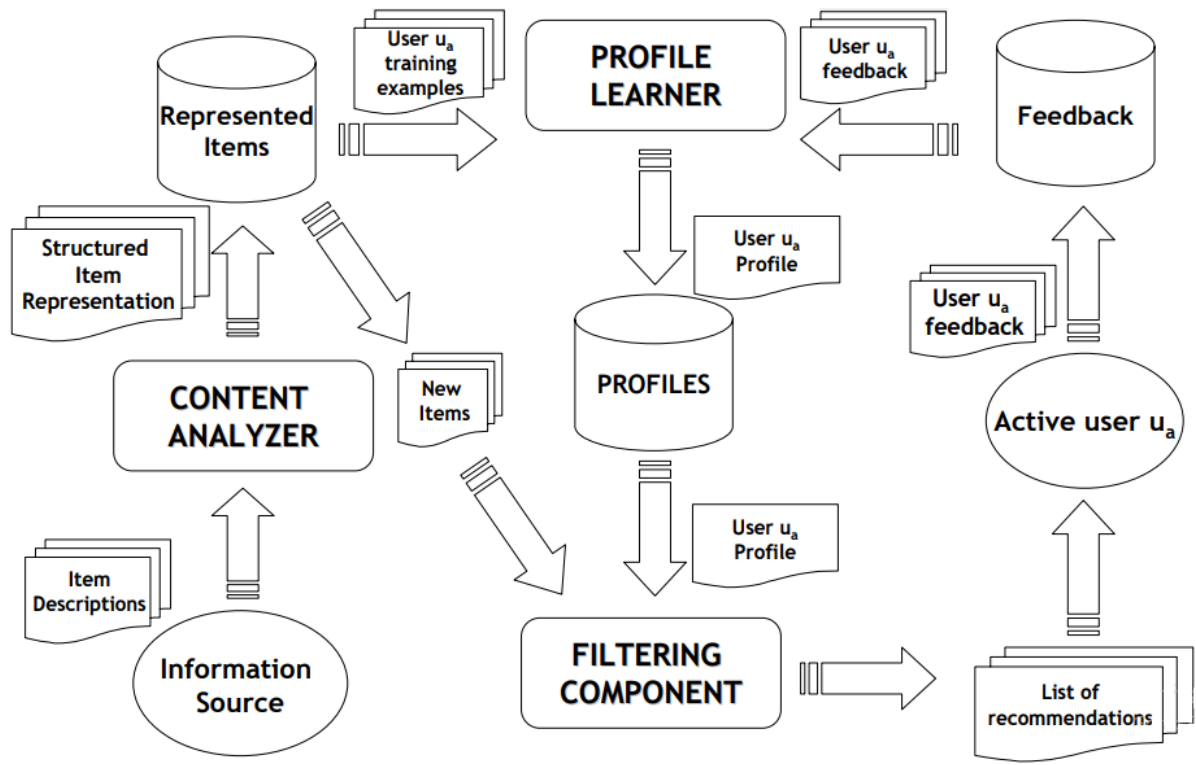

*Figure 2: Content-based architecture diagram[ref]*

The advantage of CBF is the lack of data needed from other users in order to recommend the initial user's recommendation list. Therefore, depending on what data is needed, CBF allows transparency that provides explanation for recommended items that lists features of contents that caused an item to be recommended, especially when it yields a capacity to recommend new and unknown items, making CBF have a no first-rater problem. An example of CBF into use is with LIBRA which is a content-based book recommendation system that uses information about book gathered on the web. Its algorithm uses naïve Bayes classifier that extracts information from the web and learns from user profiles in there. The system is able to provide an explanation of each user's recommendations by listing the highest ratings of features, giving the users confidence on the recommendation provided to users by the system itself.

However, automatic assignment of features to items might be insufficient purpose to define if the analyzed content is not liked by user. Certain drawbacks in CBF may include to certain users where the user is going to be recommended similar items that has already been rated. Another drawback may be the overspecialization and obviousness in recommendation to the user (e.g. 'Star Trek' being suggested to a science-fiction fan – accurate but not useful), making users not wanting algorithms that produce better ratings but rather sensible recommendations. Thus, the major disadvantage of CBF technique is the need of an in-depth knowledge and description of the features of the items in profile.

## 2.3 Collaborative Filtering

Collaborative filtering is used mostly when metadata cannot be easily described such as movies and music, i.e. the idea of collaborative filtering is in finding what the users in a community share appreciations. Its technique works by building a database of preferences for items by users – user-item matrix technique (see Figure 3). Next, it matches similarities of user profiles based on relevant information of interest. To be more exact, the user gets recommendations to specific items that have not been rated before but were positively rated by other users with similar interests. Henceforth, RS that are produced by CF can be of either prediction or recommendation itself. Such potential of similarity between users are named 'neighbors'. Figure 3 explains further the prediction of items toward the users.

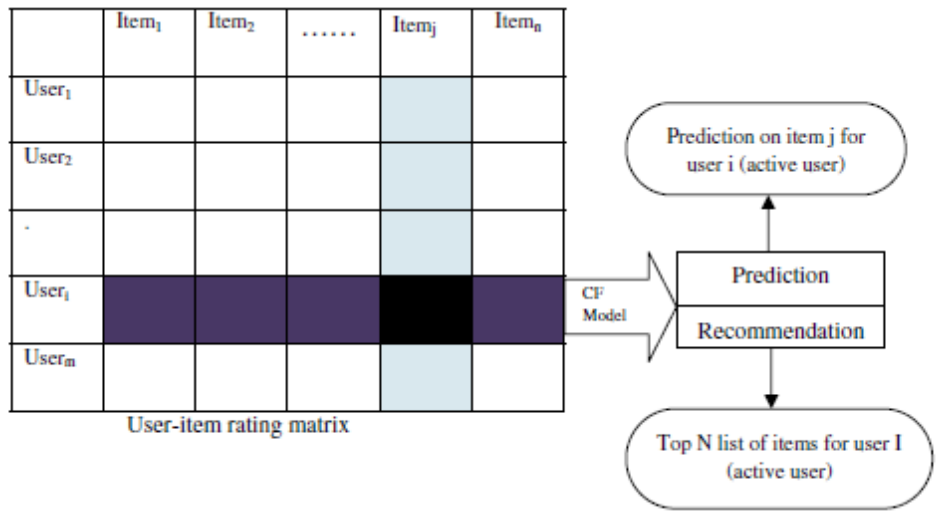

*Figure 3: Top N items the user will like the most. Collaborative filtering[6]*

Herlocker et al.[6] came with a new automated collaborative filtering (hereafter: ACF) systems that predicts a user's similarity for items or information. So, unlike traditional content-based information filtering techniques (CBF), filtering is based on human and not machine, in which it primarily does not depend on the machine analysis for content. It filters any type of content, be it text, art work, music etc. ACF does not compete with content-based filtering though. To truly know the collaboration of users, they are more likely to trust a recommendation when they know the reasons behind that recommendation, so an explanation is really needed for users to understand the process of ACF, knowing their strength and weaknesses [6]. So building the explanation facility into recommendation to users can benefit the user in many ways. It provides a transparency and benefits psychologically speaking, such as:

- Justification where user is understanding the reason behind the recommendation so it may decide how much trust it can be put within that recommendation;
- User involvement which allows the user to add knowledge skills to complete decision processes;
- Education of the user to the processes used in generating a recommendation, so the user may better understand around the strengths and limitations.

To further understand the concepts of ACF, user first enters the profile of ratings which processes information and collects a set amount of information from the user, e.g. page-views and numeric ratings. So an explanation might need to be explained onto what kinds of preference information were used in a given explanation. In other words, what kinds of data do the profile consists of? Perhaps the user has not rated enough set of movies to allow the ACF system to provide accurate recommendations with enough assurance. So what the ACF does is as in aforementioned explanation of neighborhood, to let the ACF system locate people with similar profiles. Hence the process that is used to locate other people with similar profiles is one step closer to success of the collaborative filtering technique. As in the figure shown above, neighbors selected by the system - usually visually formed in matrices – are the best predictors for the user's current information needed; actively searching for new information will result recommendation to work in the best possible performance; however, the selected neighbors of the users may not result in the best recommendation for the said user but they are rather the most similar profiles. The similarity metric is used as a result to judge the potential neighbors. The final step in this regards, the neighbors' ratings are combined to form recommendations. According to Herlocker et al., users can detect instances in some cases where the prediction is based on a small amount of data, investigating further to determine if a recommendation is an error or an inactive item.

Although Herlocker et al.[6] study with ACF disbelieves in content-based filtering and machine learning techniques, they unfortunately failed to provide more mathematical equations behind the concepts and to truly go in-depth as to how CF works memory-based and model-based techniques. It is indispensable not to only mention the human-interaction aspect of collaboration filtering techniques of recommendation, but also onto the machine-learning techniques with different set of algorithms, as they aid to categorize user within its neighbors. Consequently, collaboration filtering technique can be divided into two parts: Memory-based and model-based techniques.

## 2.3.1 Memory-based technique of Collaborative Filtering

The items that have already been found play a relevant role in neighbor that shares appreciation with the user. When a neighbor of a user is found, different algorithms will play an important role in it, in which it has proven to be a success. Therefore, memory-based CF can be achieved in user-based and item-based techniques.

User-based collaborative filtering technique makes sure to check the similarities between users that compare their rating on the same item and with that information; it computes the predicted rating for an item by the active user as an average rating of the item with other active similar users. Item-based filtering technique bases upon the similarity of items and not users. It retrieves all items rated by an active user from the user-item matrix, whereas the prediction is made by taking an average of the active users rating the similar item. Pearson coefficient for correlation further describes the measurement which two variables linearly relate to one another [7].

$$s(a,u) = \frac{\sum_{i-1}^{n}(r_{a,i}-\overline{r_a})(r_{u,i}-\overline{r_u})}{\sqrt{\sum_{i-1}^{n}(r_{a,i}-\overline{r_a})^2}\sqrt{\sum_{i-1}^{n}(r_{u,i}-\overline{r_u})^2}} \qquad (2)$$

In the above equation, *s(a,u)* denotes that the similarity between two users *a* and *u, ra;i* is rating given to item *i* by the user *a, ra* is the mean rating given by user *a* while *n* is the total numbers of items in the user-item.

$$p(a,i) = \overline{r_a} + \frac{\sum_{i-1}^{n}(r_{u,i}-\overline{r_u}) \times s(a,u)}{\sum_{i-1}^{n} s(a,u)} \qquad (3)$$

In the following formula as seen above, this is called Cosine-based measurement, in which is mostly used in information retrieval and texts mining to compare text documents. In other words for the equations above, similarity of items is determined by the similarity of the ratings of those items by the users who have rated both items, in which such equation further explains in greater detail as follows:

$$s(\overline{u} \cdot \overline{v}) = \frac{\sum_i r_{u,i} r_{v,i}}{\sqrt{\sum_i r_{u,i}^2} \times \sqrt{\sum_i r_{v,i}^2}} \qquad (4)$$

### 2.3.2 Model-based technique of Collaborative Filtering

To further state the above-mentioned missing gap by Herlocker et al., apart from using the ACF with its neighborhood-based techniques, there is another possibility to use the machine learning technique which is called model-based technique. This technique can be done by the model building process which as stated, uses machine learning or data mining techniques. They help tremendously as they can quickly recommend a set of items by using models similar to neighborhood-based recommender systems techniques. Such examples are Dimensionality Reduction technique (SVD), Matrix Completion Technique, as well as regression and clustering.

The use of such algorithms has changed the manners of how recommendation system works. In greater detail, it went from recommendation by the active users, to recommending the question of 'when' to really consume a product. Examining a few said learning algorithms used in model-based recommender systems will be further mentioned below.

**Clustering:** As once mentioned before in the beginning section, clustering algorithm partitions a set of data into a set of 'sub-clusters' in order to discover a meaningful groups that exists within. Once formed, the opinions of users within a cluster can be averaged out and makes recommendations for individual users. Therefore, a good clustering will create a high quality of data between users. A user can have a partial participation in different set of clusters, making recommendations based on the average across the clusters of participations. Self-organizing map (SOM) is the most commonly used amongst the different types of clustering methods. The SOM uses an unsupervised learning method that is based on artificial neuron clustering technique. To further explain, it uses the input nodes which are vectors with high dimensionality, while the output nodes form a map, such in this example it has 5 by 5 layout in the figure below.

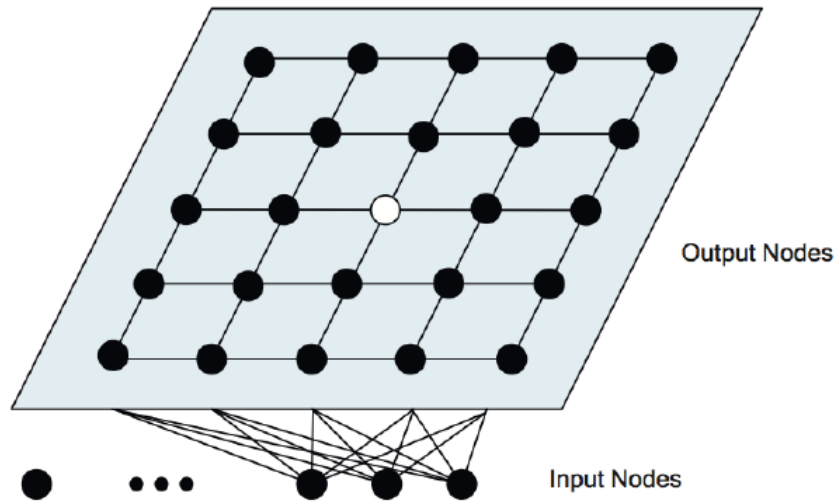

*Figure 4: SOM input nodes and output nodes on a 5 by 5 layout[ref].*

**Regression:** It is used when two or more variables are systematically connected by linear relationship. In other words, it analyzes associative relationships between dependent and its mirror independent variables. Some examples of the use of regressions contain prediction and curve fitting which can be useful to identify some sort of trending within dataset, whether be it linear, parabolic, or other forms.

**Matrix completion technique:** The essence of matrix completion technique mostly signifies to predict the unknown values within user-item matrices. K-nearest neighbor is one the techniques that is most popular amongst collaborative filtering recommendation systems. They depend upon the historical data of ratings amongst users on items – usually active users; however, can be a problem due to users not rating most of the items and represented within matrix, causing the inability to give correct and accurate recommendations.

The model-based and its machine learning techniques, they solve a sparsely problem. The drawback of the technique is due to its computationally expensive and the memory capacity usage is extraordinary intensive. The major drawback however, is the cold-state problem

To conclude it all, the collaborative filtering and its techniques have major advantages over CBF in that it has an ability to perform in domains where there is not much content to begin with or the content is difficult for a computer system to analyze. One of the advantages is that it can recommend items relevant to the users even despite not having the content in the user's profile. Even despite the successes of CF techniques, their potential problem that cannot be ignored is the cold-state problem (CS).

## 2.4 Hybrid filtering

Hybrid filtering in recommendation system combines different types of techniques in order to avoid problems and limitations of the aforementioned pure recommendation systems. The prototype behind hybrid filtering technique is to combine different set of algorithms in order to provide more accurate, yet effective recommendations as opposed to only one algorithm [8]. Using a variety of recommendation techniques will surpass the weakness behind a single technique by itself in a combined model. Therefore the main reason behind hybrid filtering is to combine collaborative filtering with content-based filtering in order to improve accuracy.

To accomplish hybrid filtering, one could categorize in various ways such as:
1. Implement collaborative and content-based methods individually by collecting their predictions;
2. Integrate characteristics between content-based and collaborative methods;
3. Consolidate a model that integrates both content-based and collaborative characteristics.

To further state the correlation between CBF and CF techniques within hybrid filtering, figure 1 shows the methods that CBF and CF recommendations estimate individually, in order to combine a better recommendation.

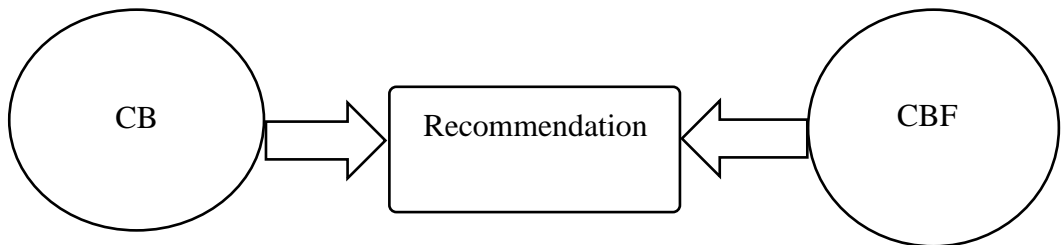

*Figure 5: CBF and CF combined to yield a better recommendation.*

Furthermore, figure 2 shows an integration of CBF characteristics into the CF approach. This especially overcomes the cold start problem in collaborative filtering and problem with overspecialization of content-based filtering:

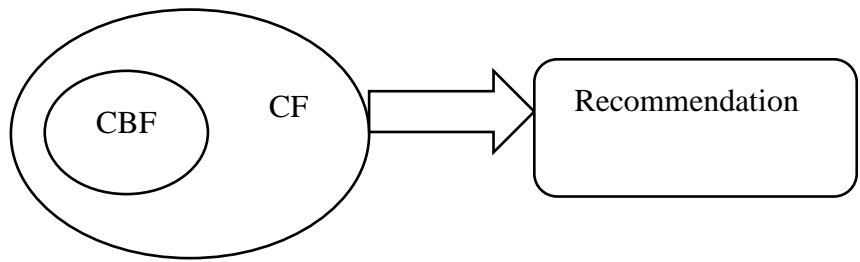

*Figure 6: CBF characteristics integrated into the CF approach.*

In the following figure shows that both CBF and CF can be combined into set of features, in a unified model in which it improves effectiveness of recommendation process:

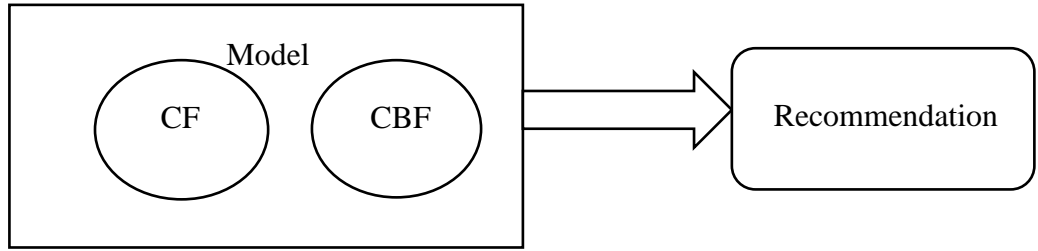
*Figure 7: CBF with CF in a constructed unifying model.*

Finally, the last figure of hybrid filtering technique shows the method that integrates CF characteristics into a CBF approach:

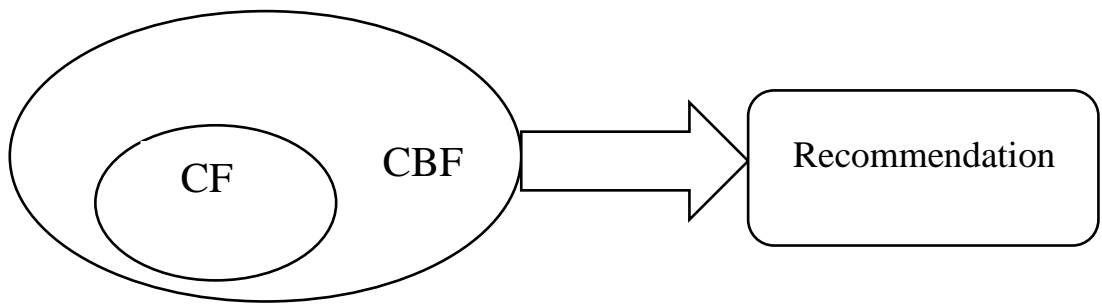
*Figure 8: CF features inside CBF approach that yields recommendation.*

Content-based filtering system has an allowance towards cold-state items in exchange to lower accuracy compared to collaborative filtering. Subsequently, collaborative filtering has a frequent rate of accurate information in exchange to cold-state problem. Hybrid approach tries to collect these different kinds of information in order for it to get an efficient recommendation result.

According to Thorat et al, Burke presented taxonomy for hybrid recommendation systems can be classified into seven following classes [8]:

**Weighted:** Combines the results of different recommenders to generate a recommendation list by using a linear formula to integrate scores from each of the techniques. An example is with P-tango whereas the system consists content-based and collaborative filtering. Initially, they are given the same weight but do fluctuate when predictions get confirmed or otherwise.

**Switching Hybridization:** The switching hybrid has the ability to completely avoid specific to one method e.g. the new user problem of content-based recommender, switching it to a collaborative recommendation system. The system is sensitive when it comes to strengths and weaknesses to recommenders. The main disadvantage of switching hybridization is its featuring of complexities to recommendation systems. DailyLearner uses both content-based and collaborative hybrid where the content-based is first engaged before collaborative filtering in situations where content-based system cannot make a recommendation with sufficient evidence.

**Cascade:** The cascade hybridization technique applies an improvement in terms of constructing an order of preferences amongst items. The default technique is refined by another recommendation technique, i.e. the first recommendation technique outputs a list of recommendations and in turn improved by another recommendation technique.

**Mixed:** It combines recommendation results at the same time rather than having one recommendation per item which in return, they will be introduced together. This is based on merging and presenting multiple rated lists into a single one. An example would be with PTV system that recommends a TV viewing schedule for the user that combines content-based together with collaborative systems to form a set of said schedule for the user.

**Feature combination:** A specified recommendation technique is sent to another recommendation technique. So the working of the recommender depends on the data that has an availability to modify by the contributing recommendation technique. Such example is with Pipper that uses collaborative filter's rating in content-based system as a feature – e.g. rating – and it does not rely on collaborative data itself.

**Feature augmentation:** Similar to feature combination but contributor instead gives more unique characteristics, making it more flexible. It also requires additional performance and its functionality from RS. Feature augmentation is superior to its neighbor feature combination due to number of features it offers to the primary recommender.

**Meta-level:** When it comes to sparsity problem, meta-level is able to solve it with ease on collaborative filtering technique by using the model of the first technique as input for the second one, i.e. the first technique substitutes the data for the second one.

## 2.5 Machine Learning in Recommender Systems

Recommendation systems in today's era of time, has been tremendously used within e-commerce websites, specifically Netflix and Amazon to name a few; however, according to Cowan et al. [9], the field of RS has its origins created since 1992 with the introduction of a new RS named Tapestry. After further practices within RS field, researchers studied the use of algorithms from machine learning (ML), an area of artificial intelligence (AI). Machine learning has been studied since the late 1950s [9] and as a result, there is a plethora of ML algorithms (e.g. k-nearest neighbor, clustering, Bayes' network). As mentioned above, ML is being used to provide a better use for RS, although ML's algorithms become more difficult and confusing for fitting within RS, making it a challenge to tackle the use of ML algorithms in RSs.

The number or choices and variations for data scientists make it challenging to deal with ML algorithms. Therefore, one must ask questions and have two main goals:
1. Identify which ML algorithms are most commonly used in RS;
2. To question openly about RS development that might be impacted by data science.

According to the author, 26 publications for ML were retained and ready to be analyzed, namely from books, conferences, and patents. The results and conclusions from aforementioned publications are presented in Table 2 for the most used ML algorithms when it comes to recommendation systems [9].

Table 2: Types of Machine Learning algorithms used in recommender systems.

| Category | Total |
|---|---|
| Bayesian | 7 |
| Decision Tree | 5 |
| Matrix factorization-based | 4 |
| Neighbor-based | 4 |
| Neural Network | 4 |

As shown in Table 2, Bayesian is the most used ML algorithm when it comes for recommendation systems, giving 7 out of 26 publications as stated by the author. Both Bayesian and Decision Trees have similarities when it comes to calculations, as both are a popular choice.

**2.6 The use of Recommender Systems in web applications**

Recommendation algorithms are mostly known for their use in e-commerce Web sites, where they prioritize the customer's interests to generate a recommendation of items. Nowadays, plenty of applications prefer to use the items of customer's purchases and explicitly rate the customer's interests, although there can be used a variety of different attributes, namely items viewed, subject interests, or even their favorite artists.

E-commerce applications and their recommendation algorithms therefore, usually work in a rather challenging environment, such as:
1. A large retailer (e.g. Amazon, Netflix) often have large amount of datas, millions of customers with plethora amounts of catalogue items.
2. As a result, many applications require for these large retailers to generate results in realtime, no more than a split second, all while giving the most quality of recommendations.
3. New customers stereotypically have low amount of information from the products they purchased, reviewed, and/or rated (i.e. cold-state situation).
4. Customer's data is quite volatile to the point where the recommendation algorithms must respond in an immediate action of generating new information since each interaction generates customer data.

Table 3 shown for all the known system examples used within RS web applications that are well-known amongst the RS:

Table 3: System examples within RS and their product goals with models used.

| System | Product goal | Models used | Review & facts | Languages used |
|---|---|---|---|---|
| Amazon | Books and other products | <ul><li>Collaborative filtering (item-to-item based)</li><li>Search-Based Methods</li><li>Cluster Models</li></ul> | <ul><li>Pioneer of RS</li><li>Virtually sells all kinds of products.</li><li>Recommendations are provided to users on the main Web page.</li><li>Purchase/browsing behavior can be viewed as implicit rating, as opposed to explicit rating.</li></ul> | Java, Javascript, Ruby, Python, Perl |
| Netflix | DVDs, Streaming Videos | <ul><li>Hybrid approaches</li><li>Neural Networks</li><li>Restricted Boltzmann Machines</li><li>Similarity based on user and item</li></ul> | <ul><li>Founded as a mail-order Dvd, eventually to streaming.</li><li>User actions watching different items stored.</li><li>Dataset consists of 100 Mio. entries.</li><li>Quadruples of *<movie, user, rating, date>*</li></ul> | Java, Scala, Javascript. |
| Facebook | Friends, Advertisements | <ul><li>Collaborative Filtering</li><li>Matrix Factorization</li><li>Main solution: Hybrid Approach</li></ul> | <ul><li>A massive social networking website</li><li>Uses friend recommendations as one of the tools.</li><li>Uses structural relationships rather than ratings data.</li></ul> | PHP, C++, Python, Java, Perl |
| IMDb | Movies | <ul><li>Matrix Factorization</li><li>User-based collaborative filtering</li></ul> | <ul><li>Widely known and popular online movie platform.</li><li>User ratings are complemented with the Movie Tweetings.</li></ul> | PHP, SQL, ASP.net |

## 3. An in-depth analysis of cold state problem

Oftentimes, the recommender systems are purely based upon the past or present user ratings for said item, item descriptions using specific keywords within, and user profiles who share in front of other audience; however, if the information is not readily available for new users or items, the recommender system will run into a so-called *cold-start problem*, i.e. cold state problem. When it approaches in such state, the system does not know what to recommend, until another user or with enough information, will 'warm up', i.e. until it reaches its threshold of information to start producing recommendations. As an example: Which product should it be recommended to someone who visits Amazon for the first time?

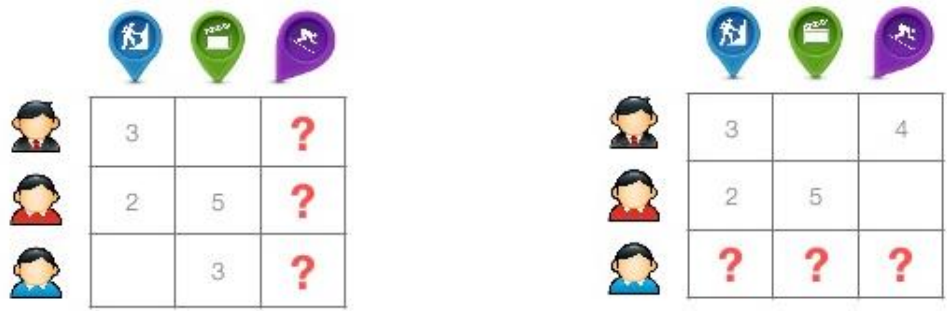

*Figure 5. Cold-state problems (On the left is 'New Item Problem'. To the right, 'New User Problem')*

To further identify, Figure 5 illustrates two sets of cold-state problem: New item problem is when a new item is added to the specific catalogue and since none has rated the item, it will never be recommended. Another one being new user problem, whereas a new user has no ratings, making it hard to predict the said ratings.

As a result, there are several approaches to possibly deal with cold-state problem, namely utilizing the baselines for cold users, extracting ratings from new users, a system to combine collaborative filtering with content-based recommenders, or even exploiting the users' social networks. With the aforementioned approaches, it is a temporary fix for information to be gathered and apply to its recommender system, especially in a situation when there are no 'warm users' or its period is short-lived; however, in many e-commerce applications, users or items may remain cold for a long, continuous amount of time or they can get back to a cold state, leading onto a *continuous cold-state (CCS)*. Such an example can be taken from Booking.com whereas the user may go for a long break and not return to a site for a while, leading to a CCS. Furthermore, a warmed, long-term user may become cold, as they change their needs over the time, wanting to desire entirely different needs over the years. Such users and cases frequently tend to happen in the aforementioned accomodation sites where the user may only give a visit for business or holiday trips which may be at best, unpredictable for a system. Therefore, a classical approach to a cold-state problem may fail in case of CCS, since they assume the user will remain warm after a visit. In this section will be further discussed about the CCS (continuous cold-state) and a thorough way to give a new user some incentive reasons to rate through Hossein et al. Ask-to-rate technique which divides in non-adaptive and adaptive methods.

### 3.1 User Continuous Cold-State

Kiseleva et al.[10] stated that there are different ways to define the users who become inactive:

1. **Classical CS:** This is the case of a new or rare users.
2. **Incosistency:** User's interest changes over a period of time and has an incosistent interests over different times.
3. **Identification:** User's identity may act differently, failing to match the same data from the same user.

The aforementioned cases originate commonly with e-commerce websites. The user may appear new from a different device, or even another identity logs into the same user, meaning the system may have a failure of identification matching. Henceforth, some users change their interests over a period of time, e.g. the movie preferences evolve and changes onto a different movie niche, user's travelling has changed in which it tends to happen frequently or in the case of e-mail marketing, the user becomes inactive and tend to not open, nor click them. All of such problems arise for collaborative filtering, as well as content-based approaches, due to user simply not engaging in user ratings, activities and profiles being empty.

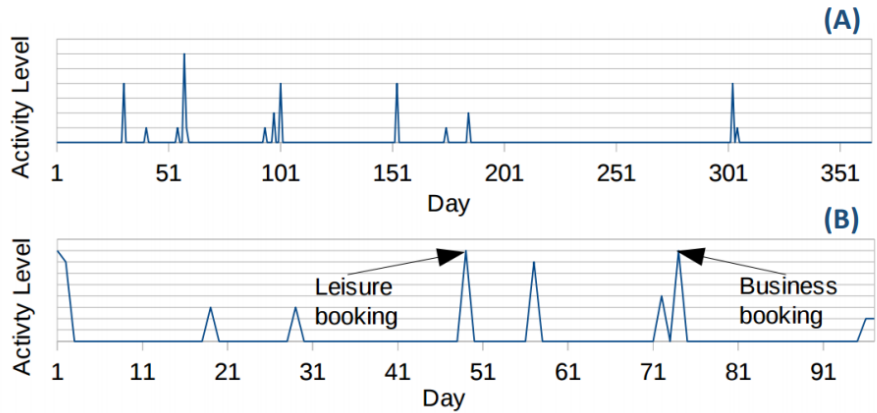

*Figure 6: CCS users at Booking.com. Activity levels of two randomly chosen users over time. (A) Signifies user having an occasional activity throughout a year. (B) Different personas within a user by making a leisure and a business booking with no activity in between. Sourced: [10].*

Figure 6 (A) shows new users arriving frequently or may appear new when they do not log in or use a different device altogether. Most users change their interests over time or even change personas on shorter time. With Figure 6 (B) an example could be simply due to weather or their travel purpose, wanting to book different type of trips.

**3.2 Item Continuous Cold-State**

In a uniformal way, CCS problems also defines for items:

- **Classical CS:** This is the case of a new or rare items' clicks or impressions by users.
- **Incosistency:** Item 'trend' changes over time; item appeals to different type of users.
- **Identification:** Failure to match the same data for the item.

New items are in a constant appearance in e-commerce applications. Some of the itemes are interesting only to niche audience, e.g. books or movies specifically on the topic. Items may be unpredictable and changes over time, as a new, better version of the same article takes over the older one, e.g. a new phone releases and takes more sales than the older one. Item incosistency may derive as a context from news and trendy conversations, defining the item as a topic shift. When more items are added into the e-commerce site catalogue, or when even more catalogues are added within, it creates a problem for matching items, hence a failure of identification. Figure 7 further shows the example of Booking.com with a continuous cold-state problem within items.

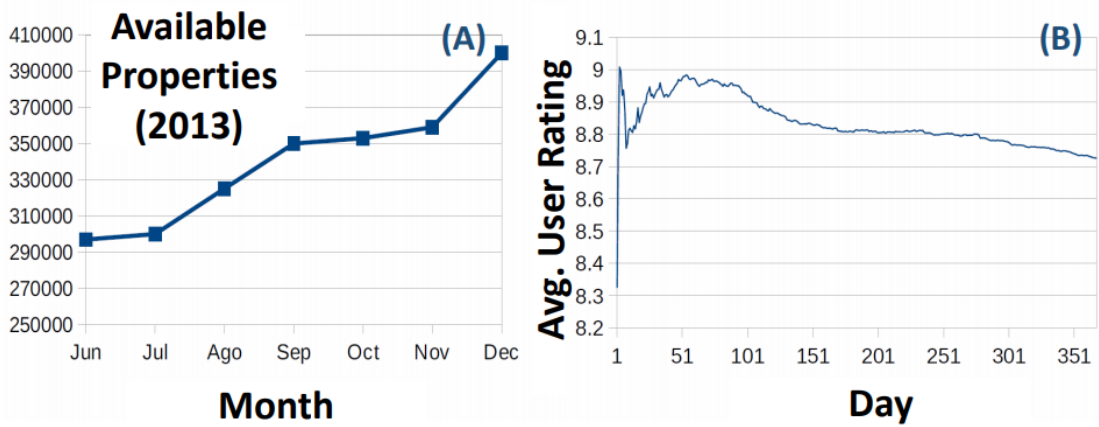

*Figure 7: CCS Items at Booking.com. (A) Signifies thousands of properties are added every month. (B) User ratings over a randomly chosen hotel change over the year. Sourced: [10].*

### 3.3 Non-adaptive methods

The non-adaptive method within ask-to-rate technique makes it possible to present all items as the same items in new users, regardless of changes within new user's knowledge.

In 2002, MovieLens researchers came up with new strategies on solving and learning about new users. The said strategies were focused on the issue of which items to be recommended to the new users; their outcome through different strategies that were used and experimented online on their MovieLens datasets, considered user effort and recommendation accuracy which is related to the user experience. The suggested methods were measured based on rating prediction accuracy and they were as follows [15]:

**Random Strategy:** It selects the items randomly which learns about new user preferences through all the available items. Random strategy is most commonly used for comparisons, e.g. for items. Unfortunately, the analysis of the rating matrix is not intelligent enough and shows it through offline and online experiments that it needs more user effort in it.

**Popularity strategy:** This takes into account on how many users have rated an item. The items are presented to the new users based on the number of ratings that they have been given by the users. Furthermore, popularity strategy has an equation: Item $a_t$ is computed, where $r_{at}$ shows rating.

$$Popularity(a_t) = |r_{at}| \qquad (5)$$

The point of the equation means to minimize user effort. Ratings however, may be uninformative since most users like popular items.

To point out more with MovieLens idea of ask-to-rate, in 2008 the researchers extended its group to improve order of items and extract opinions of new users right away at registration time. They proposed an offline simulation framework and online experiment with real users of MovieLens live RS [11, 15].

### 3.4 Adaptive Method

The items are consistent with each new user's opinions. So the items get present right away when the new user attempts to rate an item. It is more controlled more effectively than in non-adaptive approaches. Adaptive methods take approaches of user's historical ratings when it is in its initial state. To deal with the cold-state situation however, there are a few adaptive approaches such as item-item personalized, information gain through clustered neighbors (IGCN), naive Bayes, and clustering.

### 3.5 Methodology used in CS

When it comes to the aforementioned continuous cold-state situation problem, the testing environment with A/B production was taken advantage for online travel agency e-commerce sites such as Bookin.com. The A/B testing randomly splits users to see two different versions of the website – baseline or variant of it. When it comes to baseline, the Kiseleva et al. used a non-contextualized ranker corresponding to the live system [10]. This system is optimized and therefore is trained to be pressured by a massive volume of traffic. In terms of evaluation metrics, Kiseleva et al. used in the A/B testing the likes of clicks-per-user and click-through-rate (CTR). The motivation behind this task was to increase customer engagement, since it was used with an exploratory task. More clicks and CTR signifies that the user clicks more on the suggested destinations, interacting more with the system.

Another methodology used in cold state situation when it comes to movie recommendations can be an example of data used from the 'MovieLens' assembled by the 'GroupLens' project [11]. Its main purpose is to list movie ratings group by a person. Approximately, each person would rate at the very least 20 movies. All the MovieLens observations extract ratings between 1 and 5 respectively. Thus, in order to obtain accurate actor lists and director information, it heads onto the Internet Movie Database (imdb.com) by downloading pages and extracting only the top ten actors, eliminating the ones who have not appeared more than one movie.

The methodology of this MovieLens took 943 people that were made for recommendations, averaging 85 observations per person in the training set in its entirety, 331 movies into the testing set (out of 1682 of movies in total from imdb at the time) [12]. So there are three modes identified to test the application: Role of the customer's purchase of item or that a customer will both, like and purchase the product (movie), and guess the customer's rating on an item that was previously purchased. These tests will be further elaborated below [12]:

1. Implicit Rating Prediction - This refers to the prediction of data such as the purchase history itself. Therefore, this purchase cannot indispensably be a total satisfaction by the customer, but an implicit need for such purchase or desire for an item. In MovieLens data it predicts that the customer has rated a movie, being comparable to predicting a customer's purchase. Implicit rating is appropriate when explicit rating is not much available but satisfied enough to recommend products that the user is likely to purchase it.

2. Prediction of Rating itself - The purpose of this is to predict the implicit rating and items simultaneously. This can be classified into each person or a movie pair that does not occur in the observation (person as $p$ and $m$ as a movie):

    A) $p_i$ is rated as $m_j \geq 4$.
    B) $p_i$ did not rate $m_j \geq 4$.

    Condition of B) could be applied that person $i$ did not rate movie $j$.

3. Rating Imputation - Rating imputation is a prediction of ratings for items that has implicit ratings observations. Specifically, a question rises: "Given the person's seen movies $x$, how likely is it to rate it $\geq 4$?" The prior knowledge gives in the $x$ which the person may have watched, meaning that this may be an implicit rating observation. The rating imputation extracts from the MovieLens data and therefore predicts the ratings.

In the real-world applications there are data sets which implicit rating observations are available in large quantities, but the rating items is missing. Henceforth, this rating imputation fills the gap in the missed values. This is not the first time the rating imputation has been used in recommendation system: "A combination with empirical analysis of predictive algorithms collaborative filtering" to name amongst many works for recommendation system.

## 4. Discussion

To conclude this in a possible CS treatment of data, there are a few algorithms that individually handle with the case of e-commerce websites. Amazon uses topic diversification algorithms to improve its recommendations, whereas the system uses collaborative filtering method to overcome issue with scalability that generates a table of similar items in an offline mode, by using item-to-item matrix [13]. The system recommends based on the user's previous purchase history and throughout the searches the user has made an active seeking throughout the e-commerce site over the items which helps the system identify the user's recent activity. Content-based techniques matches the content to the user's search characteristics, as it normally gets based upon the user's information throughout the search, ignoring the contributions from other users as with the case of collaborative techniques.

To directly deal with CCS problem, there are traditional approaches when it comes to dealing with user continuous problem. Recommendations based on social networks is an interesting new approach as it can supplement missing information previously from the e-commerce application. As a prime example would be by using Facebook, using recommendations based on likes are proposed in, even though the identification problem within a user's device usage or multiple personnas within a user remains a trouble for now. When a user clicks throughout the e-commerce website, the browsing behaves and recommends based on what user is clicking inside the site; however, this still tends to delay the recommendation system until enough clicks have been made by the user. A more promising approach to this would be a contextual-based recommendation or as aforementioned, the content-based recommendations. To begin with a content-based recommendation therefore, has proven to be effective due to exploitage to find an initial item based on a single interaction that the user has clicked throughout the site. Context-based recommendation however, is particularly promising when it comes to solving CCS, by being based upon the context of the current visitor and the similar behaviors they posses on a similar context [14]. They define a set of features, e.g. time, location, device to pick a few from many. These datas cluster users into a context, making the context-based recommendation to also have created a gap for cold-state of a cold context that the system has never seen the behavior of that particular user before.

In summary, CF systems yield recommendation based on user-to-user similarity but the new user encounters a serious problem in the CF approach since it has to acquire some data about the new user. Hence, the non-adaptive with adaptive approaches were shown for ask-to-rate technique that exactly deals with new user problem, even though it still has its flaws with a few methods that have been proposed to deal with CS. In conclusion, there can new methods in the future that will perhaps deal more with CS problem.

Giving an example, Amazon recently created a new, more incentive ways to engage users. Such system is called 'Amazon Prime' [16] whereas the user subscribes to the Amazon website itself, giving more discounts and incentive reasons for the user to stay in a warm state.

## 5. Conclusion

Today, there are plenty of information available on the internet but it is not easy for each user to find relevant information in short amount of time. In order to overcome this problem, the recommendation system was introduced. So in this paper is presented a RS that open new opportunities to retrieve personalized information on the internet. The problems and solutions are presented in the most known recommendation system models and techniques that are analyzed in greater details. Various algorithms and machine learning are mentioned for their quality and performance with a few examples, all convenient for recommendation system. This paper also discussed two traditional (i.e. content-based and collaborative filtering) techniques for recommendation which highlighted their overall strength and performances, together with their challenges and ways to overcome.

However, the big issues unfortunately is in fact that user interests and taste changes with time – as a result, social networks are the first and most active place to be notified with such changes. Furthermore, these recommendation system techniques are limited in the way they operate – e.g. an e-commerce site can yield recommendations only within purchases made or products browsed by users. So the developers have a good chunk of data from their own monitoring their incentive e-commerce platform (e.g. Amazon prime) and social networks which increases user's activity, creating precisive and up-to-date user profiles that can be used for recommendation such as ads, consumer goods, and travel to name amongst many use cases. Another problem within recommendation system is data sparsity whereas an e-commerce site has a big catalogue but very few purchases and this especially is problematic to a new user which would be virtually impossible to recommend anything, hence why mentioning the context-aware recommendation system may help deal with the situation in a cold state. So when a user visits the website, its browsing behavior is used to estimate its intent of after a few clicks through the website – be it searching for different products or user profiles. Another good solution to the cold state situation is the use of data gathered through social since the user profile is – most of the time – already detailed.

## References

[1]. Mayer-Schönberger, V., & Cukier, K. (2013). *Big data: A revolution that will transform how we live, work, and think*. Houghton Mifflin Harcourt.

[2]. Arora, S. (2016). Recommendation engines: How Amazon and Netflix are winning the personalization battle. *MarTech Advisor*.

[3]. Isinkaye, F. O., Folajimi, Y. O., & Ojokoh, B. A. (2015). Recommendation systems: Principles, methods and evaluation. *Egyptian Informatics Journal*, *16*(3), 261-273.

[4]. Pazzani, M. J., & Billsus, D. (2007). Content-based recommendation systems. In *The adaptive web* (pp. 325-341). Springer, Berlin, Heidelberg.

[5]. Di Noia, T., & Ostuni, V. C. (2015, July). Recommender systems and linked open data. In *Reasoning Web International Summer School* (pp. 88-113). Springer, Cham.

[6]. Lops, P., De Gemmis, M., & Semeraro, G. (2011). Content-based recommender systems: State of the art and trends. In *Recommender systems handbook* (pp. 73-105). Springer, Boston, MA.

[7]. Herlocker, J. L., Konstan, J. A., & Riedl, J. (2000, December). Explaining collaborative filtering recommendations. In *Proceedings of the 2000 ACM conference on Computer supported cooperative work* (pp. 241-250).

[8]. Melville, P., Mooney, R. J., & Nagarajan, R. (2002). Content-boosted collaborative filtering for improved recommendations. *Aaai/iaai*, 23, 187-192.

[9]. Thorat, P. B., Goudar, R. M., & Barve, S. (2015). Survey on collaborative filtering, content-based filtering and hybrid recommendation system. *International Journal of Computer Applications*, *110*(4), 31-36.

[10]. Portugal, I., Alencar, P., & Cowan, D. (2018). The use of machine learning algorithms in recommender systems: A systematic review. *Expert Systems with Applications*, *97*, 205-227.

[11]. Kiseleva, J., Tuzhilin, A., Kamps, J., Mueller, M. J., Bernardi, L., Davis, C., ... & Hiemstra, D. (2016). Beyond movie recommendations: Solving the continuous cold start problem in e-commercerecommendations. *arXiv preprint arXiv:1607.07904*.

[12]. Ekstrand, M. D., Kluver, D., Harper, F. M., & Konstan, J. A. (2015, September). Letting users choose recommender algorithms: An experimental study. In *Proceedings of the 9th ACM Conference on Recommender Systems* (pp. 11-18).

[13]. Schein, A. I., Popescul, A., Ungar, L. H., & Pennock, D. M. (2002, August). Methods and metrics for cold-start recommendations. In *Proceedings of the 25th annual international ACM SIGIR conference on Research and development in information retrieval* (pp. 253-260).

[14]. Ziegler, C. N., McNee, S. M., Konstan, J. A., & Lausen, G. (2005, May). Improving recommendation lists through topic diversification. In *Proceedings of the 14th international conference on World Wide Web* (pp. 22-32).

[15]. Adomavicius, G., & Tuzhilin, A. (2011). Context-aware recommender systems. In *Recommender systems handbook* (pp. 217-253). Springer, Boston, MA.

[16]. Nadimi-Shahraki, M. H., & Bahadorpour, M. (2014). Cold-start problem in collaborative recommender systems: Efficient methods based on ask-to-rate technique. *Journal of computing and information technology*, *22*(2), 105-113.

[17]. Dunn, Jeff. *"Amazon Keeps Giving Goodies Away to Prime Members Because It Pays off in the End."* Business Insider.

[18]. Elahi, F. B. M. M. (2019). Cold Start Solutions For Recommendation Systems. IET.